\documentclass{aa}  

\usepackage{graphicx}
\usepackage{txfonts}
\usepackage{lipsum}
\usepackage{subcaption}       
\usepackage{hyperref}

\begin{document}
\makeatletter
\@ifundefined{linenumbers}{}{\let\linenumbers\relax}
\makeatother
   \title{
The role of environment on the evolution of disc galaxies density profiles}

    \subtitle{New insights from simulations and comparison to Euclid data}

\author{ M.~Mondelin
        \inst{1}
        \and
        F.~Bournaud
        \inst{1}
        \and
        J-C.~Cuillandre
        \inst{1}
        \and
        P. Hennebelle
        \inst{1}
          }

   \institute{Universit\'e Paris-Saclay, Universit\'e Paris Cit\'e, CEA, CNRS, AIM, 91191 Gif-sur-Yvette, France \\\email{maelie.mondelin@cea.fr}}

   \date{Accepted 12 June 2025}
 
\abstract{
Galactic discs are known to have exponential radial profiles in luminosity and in stellar surface density, in their bright inner regions. Nonetheless, their faint outer regions often display a break in the profile, with either a down-bending break or an up-bending break of the density profile. Recent Euclid Early Release Observations have shown that down-bending breaks are very scarce in the Perseus cluster, which was already suspected with poorer statistics in the Virgo cluster.

We use hydrodynamic simulations of disc galaxies interacting with a Perseus-like cluster. We show that Type II profiles, corresponding to down-bending disc breaks, can be rapidly eroded by the cluster tidal field on a timescale of approximately 1\,Gyr, while Type III profiles, associated with up-bending breaks, and Type I profiles, with no significant break, remain largely unaffected. Type II profiles are eroded through a combination of dynamical processes, including tidal stirring of pre-existing stars by the cluster potential, and triggering of new star formation in the outer disc. 

Overall, our simulations show that observations of disc breaks across different environments and cosmic epochs are consistent with a coherent evolutionary picture. At high redshift, observations by JWST of disc galaxies reveal early break structures formed in relatively isolated environments. At low redshift, isolated disc galaxies in field environments continue to exhibit these break features, while dense cluster environments, as observed by Euclid in the Perseus cluster, show significant alterations to these profiles.
Our findings support a scenario in which down-bending disc break profiles result primarily from internal dynamical processes -- such as disc instabilities and resonances -- during early formation phases, and are later modified by environmental effects in dense clusters. This interpretation does not require invoking additional mechanisms such as ram-pressure stripping or variations in star formation density thresholds to explain the observed evolution of down-bending breaks among disc galaxies at various redshifts and in various environments.
}

    \keywords{Galaxies: clusters: individual: Perseus, Galaxies: interactions, Galaxies: evolution,Galaxies: fundamental parameters}

\titlerunning{Environmental effects on disc galaxy density profiles}
\authorrunning{Mondelin et al.}

   \maketitle

\section{\label{sc:Intro}Introduction}
The radial structure of galactic discs holds fundamental clues to their formation and evolution. While the canonical picture depicts a universal exponential light profile in stellar discs \citep{freeman70}, deviations are commonly observed in their outskirts, where profiles often exhibit 'breaks', i.e. transitions beyond which the slope steepens or flattens. These profiles are typically classified into three types: Type~I (pure exponential throughout), Type~II ('down-bending break'), and Type~III ('up-bending break') \citep{vanderKruit1979, pohlen06, ErwinPohlenBeckman2008}. Type~II breaks, in particular, are observed even at high redshift with JWST ($z \sim 2$, \citealt{jwst-breaks}), suggesting that they may trace key processes shaping disc galaxies over cosmic time. Despite extensive theoretical and observational efforts, the physical mechanisms behind these disc profile types remain under active debate.

Down-bending Type~II breaks have been linked to two main classes of processes: internal dynamical mechanisms and star formation regulation. Internal dynamical evolution can produce such breaks through several pathways. Non-axisymmetric features, such as bars and their resonances, redistribute angular momentum, reconfiguring the disc radial structure \citep{Elmegreen2006, Laine2016}. In particular, the outer Lindblad resonance can act as a barrier for angular momentum transfer, reshaping stellar orbits. At earlier cosmic times, violent disc instabilities also contribute to this process by pushing material outwards and steepening the outer profiles \citep{BEE07}. Stellar migration, driven by interactions with spiral arms and bars, further displaces stars from their birth radii, leading to extended, complex distributions \citep{SellwoodBinney2002, Roskar2008}. Recent simulations by \citet{Ruiz-Lara2017} further demonstrate that radial redistribution of stars -- both through outwards migration and accretion -- plays a pivotal role in shaping Type-II breaks, with their formation and maintenance being tightly linked to the interplay between in-situ and ex-situ stellar populations. However, these studies primarily focus on secular evolution in isolated environments, leaving open the critical question of how such breaks evolve under external perturbations, such as those prevalent in galaxy clusters.

In addition to these dynamical effects, the regulation of star formation in the outer disc is thought to be a major driver of Type~II break formation. Star formation in disc outskirts often becomes inefficient due to gas surface densities falling below the critical threshold needed for collapse \citep{Kennicutt1989, Schaye2004}. This threshold is governed by the turbulent state of the interstellar medium (ISM), itself regulated by a combination of feedback, gravitational instabilities, and galactic shear \citep{elmegreen02, Hennebelle2011, renaud12}. As star formation fades in low-density regions, the build-up of stellar mass naturally declines, steepening the light profile and creating a break. While both dynamical and star formation processes may contribute, their relative roles remain uncertain, especially in varying galactic environments.

In contrast, Type~III ('up-bending') profiles are often interpreted as signatures of ongoing or externally induced activity in the outskirts. Internally, extended star formation in low-density environments can contribute to the excess light beyond the break \citep{GildePaz2005, Thilker2005}. Externally, minor mergers and tidal interactions can scatter stars to large radii, adding light to the outskirts and dynamically heating the outer disc \citep{Younger2007, Eliche-Moral2011, Borlaff2014, Helmi1999}. Beyond such direct dynamical effects, these tidal interactions may also induce localized compression, increasing gas densities and promoting star formation in different regions \citep{Koopmann2004}. These features suggest a composite origin for Type~III breaks, combining internal secular growth with environmental effects.

The relative importance of these mechanisms, especially in shaping Type~II profiles, can potentially be disentangled by studying disc galaxies in different environments. Observations from the Euclid Early Release Observations (ERO) programme \citep{pipelineERO} offer a new window into this question. Deep near-infrared and optical imaging of disc galaxies in and around the Perseus cluster \citep[][hereafter Paper~I]{Mondelin2025} reveals a striking lack of Type~II profiles in dense environments, alongside a high frequency of Type~I and Type~III profiles. These findings echo earlier results in the Virgo cluster \citep{Erwin2012}, but with significantly deeper imaging and improved statistics. The observed suppression of Type~II profiles suggests that environmental processes, such as tidal interactions or gas stripping, may erase or transform these structures. Conversely, the prevalence of Type~III profiles and signatures of outer disc star formation imply that cluster-related processes might also stimulate star formation or enhance stellar migration in the outskirts, offsetting some quenching effects.

In this paper, we investigate whether the observed scarcity of Type~II breaks in clusters, and the relative survival of Type~III profiles, can be reproduced through idealised simulations of disc galaxy–cluster interactions. We examine two competing hypotheses: (1) that Type~II profiles form primarily through internal secular evolution and are subsequently disrupted in clusters, or (2) that their formation critically depends on thresholds in gas density and turbulence, which may be suppressed in dense environments. We also explore the hypothesis that the same environmental effects responsible for disrupting Type\,II profiles may drive transformations from Type\,I to Type\,III profiles, thereby contributing to the relative abundance of Type~III breaks in cluster environments. While several N-body simulation studies have successfully explored the formation and evolution of disc break types through internal dynamical processes such as radial migration, bar and spiral-driven instabilities, and minor mergers \citep[e.g.,][]{Minchev2012, Clarke2017, Pfeffer2022}, these works primarily focus either on galaxies evolving in isolation, or on the properties of galaxies averaged over cosmological volumes. Their results demonstrate the crucial role of internal secular evolution and satellite interactions in shaping Type~II and Type~III profiles. Our study complements these efforts by specifically addressing the impact of dense cluster environments on disc break evolution.

To this end, we use the RAMSES adaptive mesh refinement code \citep{ramses} to simulate disc galaxies with initial Type~I, II, and III profiles -- assumed to be the result of internal processes at higher redshift -- and evolve them in both isolation and in a time-dependent tidal field representing a Perseus-like cluster. Our simulations focus on tidal effects, with dynamical friction modelled through an additional drag term to account for the static cluster potential. Ram pressure stripping is not included, and we find it unnecessary to explain the erosion of Type~II profiles in our models. We show that tidal interactions alone can erase a Type~II break within $\sim$1~Gyr, while Type~I and III profiles remain largely unaffected. This erasure results from a combination of stellar orbital diffusion and the triggering of star formation in the disc outskirts by tidal effects.

This article is organised as follows: Section~\ref{sc:model} presents the simulation setup and physical models, followed by a description of the initial conditions in Section~\ref{sc:condinit}. Section~\ref{sec:results} analyses the evolution of Type~II profiles under cluster influence, and Section~\ref{sc:results13} examines the impact on Type~III and Type~I profiles. Finally, Section~\ref{sc:Discussion} discusses our results in the context of observations and previous studies.

\section{\label{sc:model}Simulation technique and physical models}

We present non-cosmological, hydrodynamic simulations modelling the idealized evolution of a disc galaxy that initially has a Type~I, II or III profile, either in isolation or in a galaxy cluster tidal field. In the latter case, the cluster properties are chosen to be representative for the Perseus cluster in order to compare with the results presented in Paper~I. The initial conditions for the various types of galaxy discs and the orbital parameters in the cluster potential well are described in Section~\ref{sc:condinit}.

Our idealized simulations are performed with the Adaptative Mesh Refinement (AMR) code RAMSES \citep{ramses} with physical models similar to those in \citet{B14}. The coarse level spatial resolution is 80~pc and the finest spatial resolution is 6.25~pc. Each AMR cell is refined into
$2^3$ new cells if {\it i)} its gas mass is larger than 2$\times$  10$^4$~M$_\sun$,
 or {\it ii)} the local thermal Jeans length is smaller than four times the cell length,
 or {\it iii)} it contains more than 40 particles. 
 
 An artificial pressure floor is added to high-density gas, such that the Jeans length cannot drop below four time the smallest cell size. This is typically considered to avoid artificial fragmentation \citep{truelove} and can be considered as a sub-grid pressure term accounting for small-scale turbulent motions \citep{ceverino12}. This pressure floor is similar to that in \citet{teyssier10}: 
\begin{equation}
P_{\rm floor} = 16 \, \epsilon^2 \,G \rho^2 / \gamma\pi
\end{equation}
where $G$ is the gravitational constant, $\gamma$ is the adiabatic index of the gas, $\epsilon$ the smallest cell size (1-D length), and $\rho$ the local gas density.

Thermal evolution includes atomic and fine-structure cooling assuming solar metallicity. \citet{B10} and \citet{B15} showed that this modelling approach leads to the formation of a multiphase interstellar medium (ISM), in which structures with number densities up to $10^{5-6}$\,cm$^{-3}$. The initial gas disc is assigned typical values, with a density of a few cm$^{-3}$ and a temperature of $10^4$\,K, which is the typical for such densities. Gas cooling below 100\,K is prevented to increase the computational efficiency, resolving colder temperatures would not be relevant at our numerical resolution \citep{B15}.

Star formation is modelled using a constant fraction $\epsilon_{\mathrm{SF}}$ = 5\% of gas that is converted into stars per local gravitational free-fall time, which places the modelled galaxies about on the observed Kennicutt-Schmidt relation described by \cite{Kennicutt1998} (see Sect.~\ref{subsec:sfr} and Fig.~\ref{fig:ks}). Besides, such local star formation models are known to account for the observed relations between the surface densities of gas and SFR in a large range of objects \citep[e.g.,][]{elmegreen02, renaud12, krumholz12}.

Three mechanisms for stellar feedback mechanisms are modelled:

\begin{enumerate}
\item  
Photo-ionization in \ion{H}{ii}  regions: for any stellar particle younger than 10~Myr, a photo-ionized region is computed using the Str\"omgren sphere approximation implemented by \citet{renaud13}. The spherical approximation is justified by the fact that individual \ion{H}{ii}  regions around individual OB stars are only resolved by a few cells in general, but large non-spherical \ion{H}{ii}  regions can arise from the overlap of ionized regions around clusters of young stars.
\item
Radiation pressure from young stars is implemented using the \citet{renaud13} model. The momentum available in photons is computed for a standard initial mass function, and we rely on the assumption that most of the momentum is carried by ionizing photons: hence, the momentum is acquired by the gas that is ionized, i.e. the gas contained in the \ion{H}{ii}  regions computed as described above. This allows us to determine, from physical considerations, how the available momentum $m \times v$ is distributed into a mass $m$ of gas pushed at a velocity $v$, rather than a mass $m/x$ pushed at a velocity $v\times x$, where x>1 is a dimensionless factor representing the increase in velocity relative to v. This differs from most models where the wind velocity is set to be of the order of the local escape velocity from the galaxy \citep[e.g.,][]{genel12, hopkins12}. We use a trapping parameter $\kappa = 5$ to account for multiple scattering (see \citet{renaud13} for details), such a factor is rather high but is realistic as it compensates for other sources of momentum that are not explicitly included, such as (proto)-stellar winds (e.g., \citet{krumholz-thom12, DK13}).
\item 
To include the effects of non-thermal processes, we assume that the energy injected by supernovae dissipates over a timescale of 2~Myr \citep{HF}.  

This is qualitatively equivalent to the ''delayed cooling'' approach proposed by other authors to model the blastwave phase of supernovae explosions \citep[e.g.,][]{stinson09}.
\end{enumerate}

\section{\label{sc:condinit}Initial conditions and galaxy-cluster interaction model}

\subsection{Disc galaxy initial conditions}

Our simulations model a galaxy that initially has a stable Type~I, II or III profile and enters a Perseus-like gravitational potential well. The initial disc galaxy has a baryonic mass of $10^{10.2}$\,M$_{\odot}$, purely stellar ('dry') or including a relatively modest fraction of gas ('wet', with 13\% of the baryons being gas), about representative of our observed sample from Paper~I.

The Type~II galaxy model comes from the result of a high-redshift galaxy simulation (z$\sim$1) that has formed a Type~II, double exponential profile through dynamical evolution and disc instabilities. Namely, the Type~II initial conditions are directly taken from the final stage of the simulation denoted 'run~7' in \citep[][hereafter BEE07]{BEE07}. All simulations of high-redshift unstable discs evolving into regular spirals in BEE07 end up as Type~II discs; we reused 'run~7' simply because it shows the largest change of slope at the break radius in its radial density profile. Its qualitative evolution from a violently unstable disc to a regular spiral is similar to that of 'run~1' shown in detail in BEE07 (see their Figs.~4 and 6).

We rescale the mass of this galaxy to $2\times10^{10}\,\mathrm{M}_\odot$ rather than the $5.7\times10^{10}\,\mathrm{M}_\odot$ of the BEE07 'run~7' snapshot, in order to better match the observed sample from Paper~I. In order to keep the system approximately at equilibrium and to crudely follow observed galaxy scaling relations \citep[e.g.,][]{Mo1998}, when the mass is scaled by a factor $\alpha = 2.0/5.7$, we apply a factor $\alpha^{1.2}$ to distances (to maintain constant central surface density), and a factor $\alpha^{-3/4}$ to velocities (following virial scaling). This rescaling is applied uniformly to all components: stars, gas, and dark matter.

\begin{figure*}[ht!]
\centering
\includegraphics[width=0.45\textwidth]{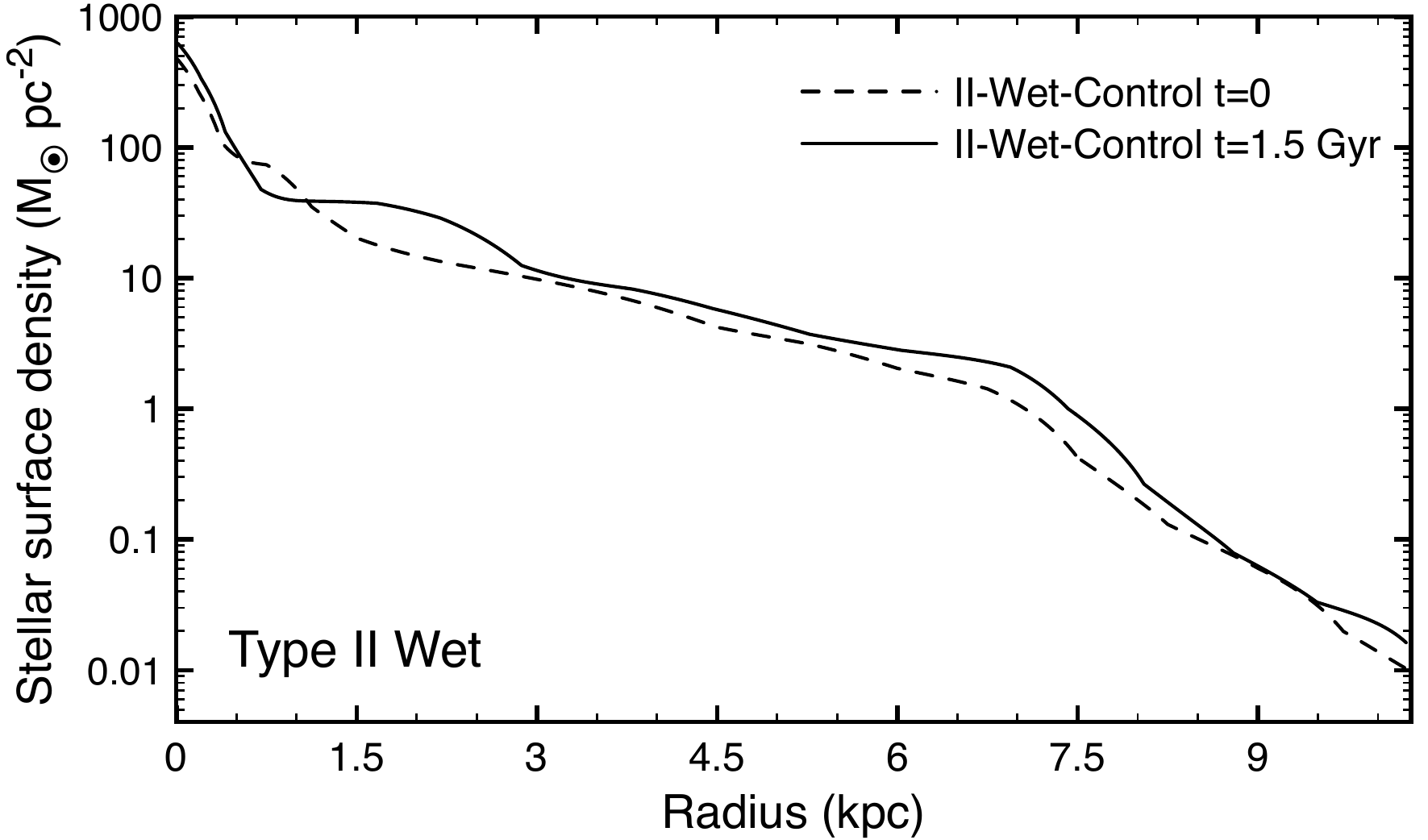}
\includegraphics[width=0.45\textwidth]{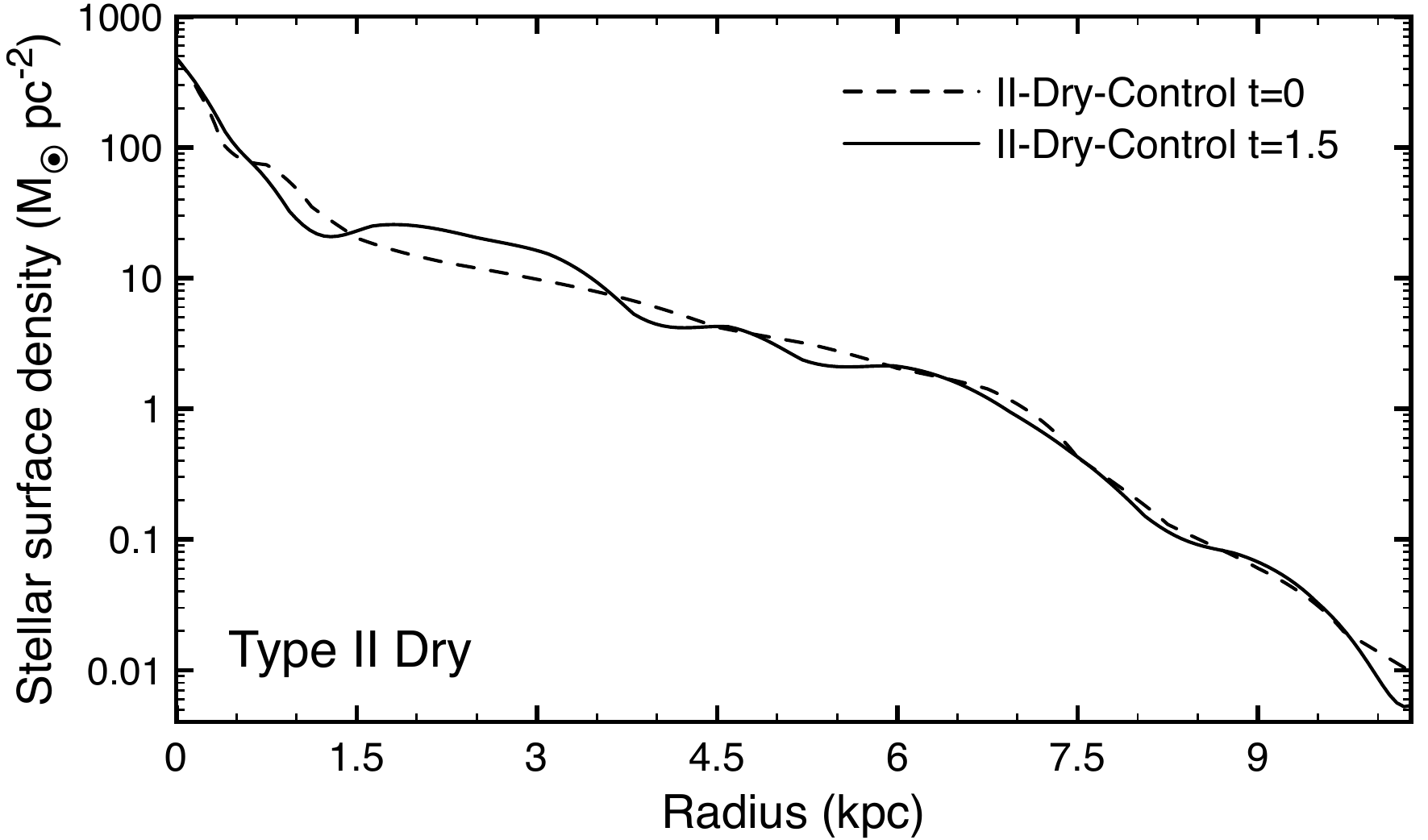}
\includegraphics[width=0.45\textwidth]{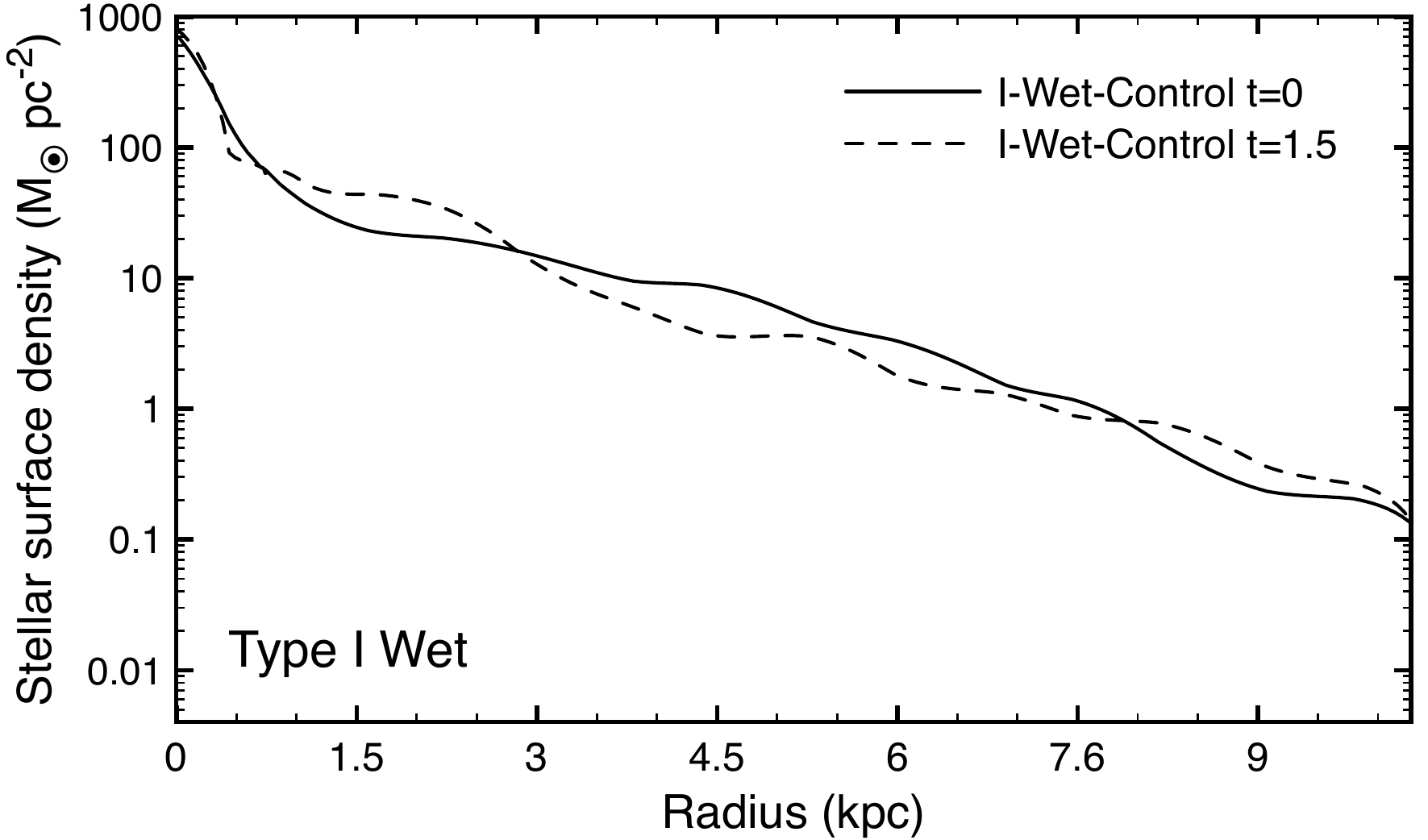}
\includegraphics[width=0.45\textwidth]{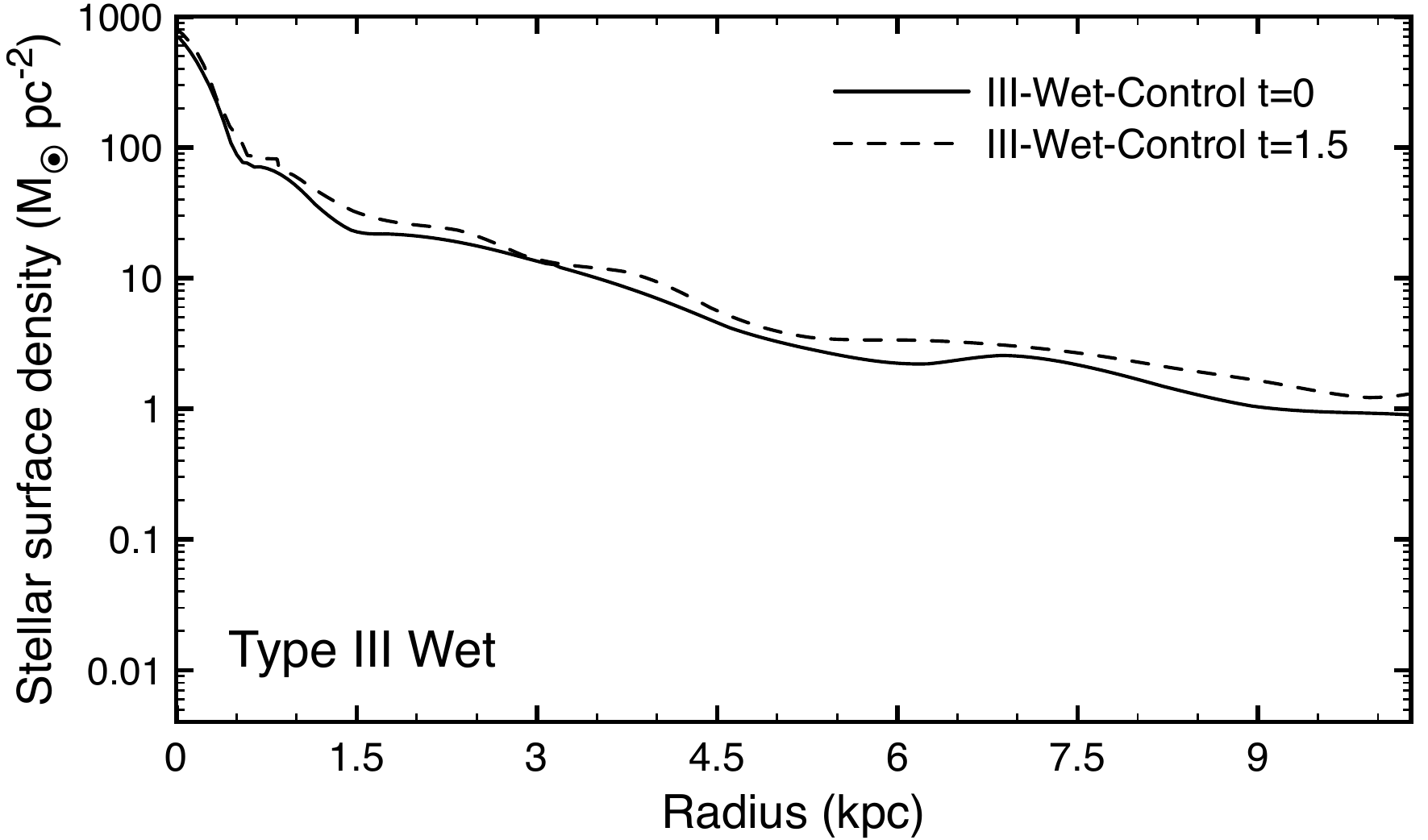}
\caption{Radial profiles of the stellar mass density in the four sets of initial conditions (solid lines), and similar profiles after 1.5\,Gyr of isolated evolution (dashed lines). }
\label{fig:ini_prof}
\end{figure*}

While this rescaling is somewhat arbitrary, it provides stable initial conditions that retain the original radial structure (see Sect.~\ref{sec:4.1}). This ensures that the Type~II down-bending break is representative of the result of disc instabilities and secular processes, such as bars, clumps, and high-redshift disc turbulence. In these initial conditions, the bulge represents 21\% of the stellar mass and the gas makes up 13\% of the total baryonic mass (17\% of the disc mass). The initial stellar surface density profile is shown in Fig.~\ref{fig:ini_prof}.

A key result from both BEE07 and our own simulations is that once the break is formed, its radius remains about constant. Although the inner and outer exponential slopes may evolve due to secular processes, the radius at which the break occurs does not significantly change. This supports the robustness of our approach to use the break radius as a structural anchor in creating Type~III and Type~I profiles.

The Type~II initial disc has a main (inner) exponential scale-length of 2.79\,kpc, a break radius at 6.78\,kpc, and an outer (steeper) scale-length of 0.83\,kpc. We also construct 'dry' Type~II initial conditions by simply converting the gas into stars; this version preserves the Type~II profile for at least 1.5\,Gyr (see Fig.~\ref{fig:ini_prof}), just like the 'wet' model.

To create Type~III initial conditions, we redistribute a fraction of the inner-disc particles into the outer disc (beyond the break radius), in such a way that the inner slope remains unchanged. The transferred stars populate the region between the break radius and three times that value, consistent with the observed extent of Type~III discs \citep[e.g.,][]{Borlaff2014, Laine2016}. We impose an outer scale-length three times larger than that of the main inner disc. The position of the break radius is explicitly preserved, consistent with the stability seen in simulations. \footnote{The forced outer profile is stopped at three times the initial break radius. Beyond that, the stellar density is negligible -- corresponding to surface brightness below 28–29\,mag\,arcsec$^{-2}$, a level comparable to that of the intra-cluster light and sky noise in the Euclid Perseus data.}

We then recompute the gravitational potential and adjust the velocities of all components by assuming that velocity scales with the square root of the local potential. While this method is approximate, it ensures dynamical consistency and approximate equilibrium. The resulting Type~III profile remains stable over at least 1.5\,Gyr of evolution in isolation  (lower-right panel of Fig.~\ref{fig:ini_prof}).

We use the same technique to create Type~I profiles, by extending the inner exponential slope uniformly out to 3 times the break radius, again maintaining the original break radius position. The stability of the resulting Type~I discs is shown in the lower-left panel of Fig.~\ref{fig:ini_prof}. Note that the profiles are shown out to similar radii in all panels to allow a visual comparison of the slope variations. In particular, the two Type\,I profiles maintain a constant slope down to stellar surface densities as low as 0.01 M$_\odot$ pc$^{-2}$.

\subsection{Cluster model and galaxy orbit}

The cluster is modelled as a gravitational potential well without baryonic or direct hydrodynamical effects -- ram pressure stripping (see \ref{sc:environments}), feedback from the central AGN, etc, are not included in the present study. Following observational constraints on the Perseus galaxy cluster \citep[e.g.,][]{ettori10, meusinger20}, we adopt a potential well modelled by a Navarro–Frenk–White (NFW) profile \citep*{NFW}, with a total mass $M_c= 2.6\times10^{15}\,\mathrm{M}_\odot$, a radius $r_{200}=2.8$\,Mpc, and a concentration parameter $c=5$. These parameters are consistent with those derived for the Perseus cluster.
The NFW density profile is given by:
\[
\rho(r) = \frac{\rho_0}{u(1+u)^2}, \quad u = \frac{r}{r_s},
\]
where \( r_s \) is the scale radius. The scale radius is defined as \( r_s = r_{\text{vir}} / c \), where \( c \) is the concentration parameter and \( r_{\text{vir}} \) is the virial radius of the halo.  
In our case, we adopt \( r_{\text{vir}} = r_{200} \), where \( r_{200} \) is the radius within which the mean density is 200 times the critical density of the Universe. This leads to a scale radius of \( r_s = 560\,\text{kpc} \).  
The total mass within the virial radius is \( M_{\text{vir}} = M_c \). Using the relation for the total mass of an NFW profile,
\[
M_{\text{vir}} = \int_0^{r_{\text{vir}}} 4\pi r^2 \rho(r) \, dr = 4\pi \rho_0 r_s^3 \left[\ln(1+c) - \frac{c}{1+c}\right],
\]
we solve for the central density \( \rho_0 \). Substituting the given parameters, we find:
\[
\rho_0 \approx 1.2 \times 10^6 \, M_\odot\,\text{kpc}^{-3}.
\]
The tidal field associated with this potential well is then added to the gravitational forces computed by the RAMSES simulations of the target disc galaxy, following the galaxy’s orbit into and around the cluster.

\begin{table*}
\centering
\begin{tabular}{lccccccccc}
\hline\hline
Run & & \multicolumn{2}{c}{Initial disc \& SF model} & & \multicolumn{3}{c}{Cluster/Orbit} & & Length of simulation \\ 
ID & & Profile & Fraction of gas & & $i$ & $r_{\mathrm{peri}}/r_{200}$ & $\tau$ & & Duration \\
\hline

II-Wet-Control  & & Type~II & 17\% & & \multicolumn{3}{c}{No cluster} & & $\sim$2.5~Gyr \\
II-Wet-Orb1     & & Type~II & 17\% & & 0°   & 0.33 & 1.0~Gyr & & $\sim$2.5~Gyr \\
II-Wet-Orb2     & & Type~II & 17\% & & 180° & 0.33 & 1.0~Gyr & & $\sim$2.5~Gyr \\
II-Wet-Orb3     & & Type~II & 17\% & & 30°  & 0.25 & 0.7~Gyr & & $\sim$1.6~Gyr \\
II-Wet-Orb4     & & Type~II & 17\% & & 150° & 0.25 & 0.7~Gyr & & $\sim$1.6~Gyr \\
II-Wet-Orb5     & & Type~II & 17\% & & 60°  & 0.50 & 1.5~Gyr & & $\sim$1.6~Gyr \\
II-Wet-Orb6     & & Type~II & 17\% & & 120° & 0.50 & 1.5~Gyr & & $\sim$1.6~Gyr \\

II-Dry-Control  & & Type~II & 0\%  & & \multicolumn{3}{c}{No cluster} & & $\sim$2.5~Gyr \\
II-Dry-Orb1     & & Type~II & 0\%  & & 0°   & 0.33 & 1.0~Gyr & & $\sim$2.5~Gyr \\
II-Dry-Orb2     & & Type~II & 0\%  & & 180° & 0.33 & 1.0~Gyr & & $\sim$2.5~Gyr \\

III-Wet-Control & & Type~III & 17\% & & \multicolumn{3}{c}{No cluster} & & <~2.0~Gyr \\
III-Wet-Orb1    & & Type~III & 17\% & & 0°   & 0.33 & 1.0~Gyr & & $\sim$1.6~Gyr \\
III-Wet-Orb2    & & Type~III & 17\% & & 180° & 0.33 & 1.0~Gyr & & $\sim$1.6~Gyr \\

I-Wet-Control   & & Type~I & 17\% & & \multicolumn{3}{c}{No cluster} & & $\sim$1.6~Gyr \\
I-Wet-Orb1      & & Type~I & 17\% & & 0°   & 0.33 & 1.0~Gyr & & $\sim$1.6~Gyr \\
I-Wet-Orb2      & & Type~I & 17\% & & 180° & 0.33 & 1.0~Gyr & & $\sim$1.6~Gyr \\

\hline
\end{tabular}
\caption{Simulation parameters. Note that the fraction of gas is the mass fraction of gas in the disc, not counting the stellar bulge component.}
\label{tab:run}
\end{table*}

Several orbits are simulated, as listed in Table~\ref{tab:run}. The initial orbits are parabolic, i.e. we assume the simulated galaxy comes from a large distance with a velocity negligible with respect to the cluster virial velocity. We vary the inclination \(i\) between the disc spin axis and the orbital angular momentum axis, where \(i = 0^\circ\) corresponds to a direct (prograde) orbit and \(i = 180^\circ\) corresponds to a retrograde orbit. For each disc Type, with or without gas, these configurations are labelled Orb1 and Orb2, respectively. We also vary the ratio between the pericentre distance and the cluster \(r_{200}\) radius, from 0.25 to 0.50. Since the cluster is modelled as a rigid potential well, dynamical friction is not self-consistently resolved in the simulation. To approximate its effects on galaxies as they move through the cluster environment, we introduce a drag force that mimics the influence of dynamical friction. This correction enhances the physical realism of the model, as without it, galaxies would escape the cluster. The drag force alters the galaxies trajectories, enabling a more accurate representation of their interactions with the intracluster medium and the tidal forces exerted by the cluster. The drag force is defined as:

$$ \vec{f} = - \frac{1}{\tau} {\left(\frac{\vec{v}}{200\,\mathrm{km}\,\mathrm{s}^{-1}}\right)}^{-2} \frac{\rho}{\rho_c} $$
where $\vec{v}$ is the velocity of the galaxy w.r.t. the cluster, $\rho$ the local cluster density (on the assumed NFW profile) and $\rho_c$ the central cluster density, and $\tau$ a drag timescale. Thus a galaxy crossing the central regions of the cluster at 200\,km\,s$^{-1}$ undergoes a drag force of timescale $\tau$. The force varies with density and velocity as expected for dynamical friction \citep{chandrasekar}. The timescale $\tau$ is chosen to be of the order of a Gyr, as found in full N-body simulations of similar cases \citep{Tormen1997, Boylan2008}.

The values of $i$, $r_{\mathrm{peri}}/r_{200}$ (where \(r_{\mathrm{peri}}\) is the pericentric distance of the galaxy orbit) and $\tau$ are given for each orbit in Table~\ref{tab:run}. On each orbit, the initial position of the galaxy w.r.t. the cluster is chosen so that the pericentre occurs 1\,Gyr after the beginning of the simulation, which corresponds to approximately one to one and a half crossing times.

Note that in this paper we do not attempt to systematically explore the parameter space. The simulated orbits are chosen arbitrarily to cover broad range of initial conditions, which, as we will show later, lead to consistent and robust trends.

\section{\label{sec:results}Results: Evolution of Type~II (down-bending break) profile in/near a galaxy cluster}

In the following three sections, we first examine the evolution of the overall density profiles of Type II galaxies. We then aim at disentangling the contributions of internal dynamical processes from those of star formation.

\subsection{Density profile evolution}\label{sec:4.1}

\begin{figure}[ht!]
\includegraphics[width=0.47\textwidth]{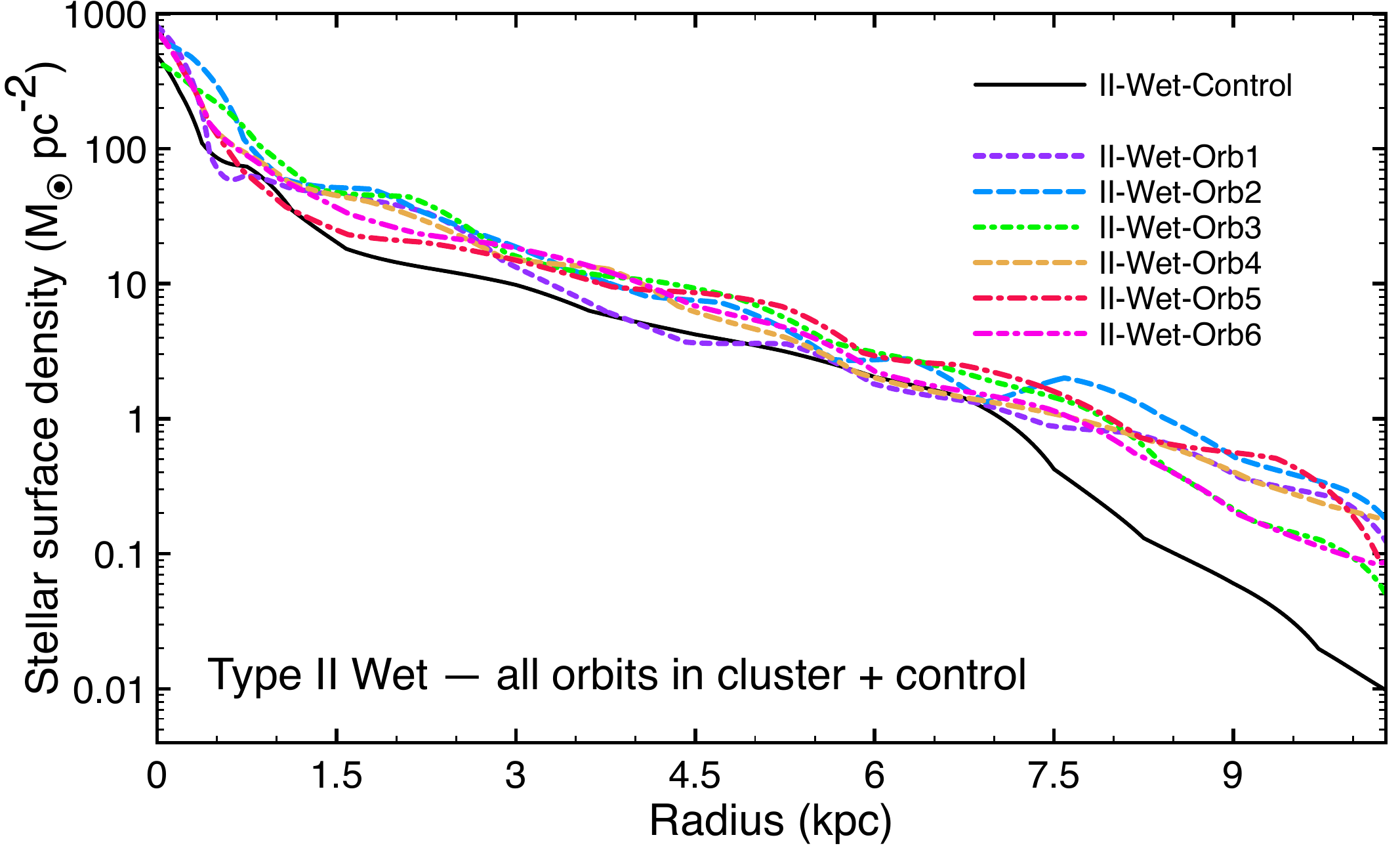}
\caption{Evolution of the Type~II, 'wet' runs, for the 6 orbits simulated in the cluster, at $t=1.5$ (compared to the isolated control run shown with a solid black line).}
\label{fig:wet_evo}
\end{figure}

\begin{figure}[ht!]
\includegraphics[width=0.47\textwidth]{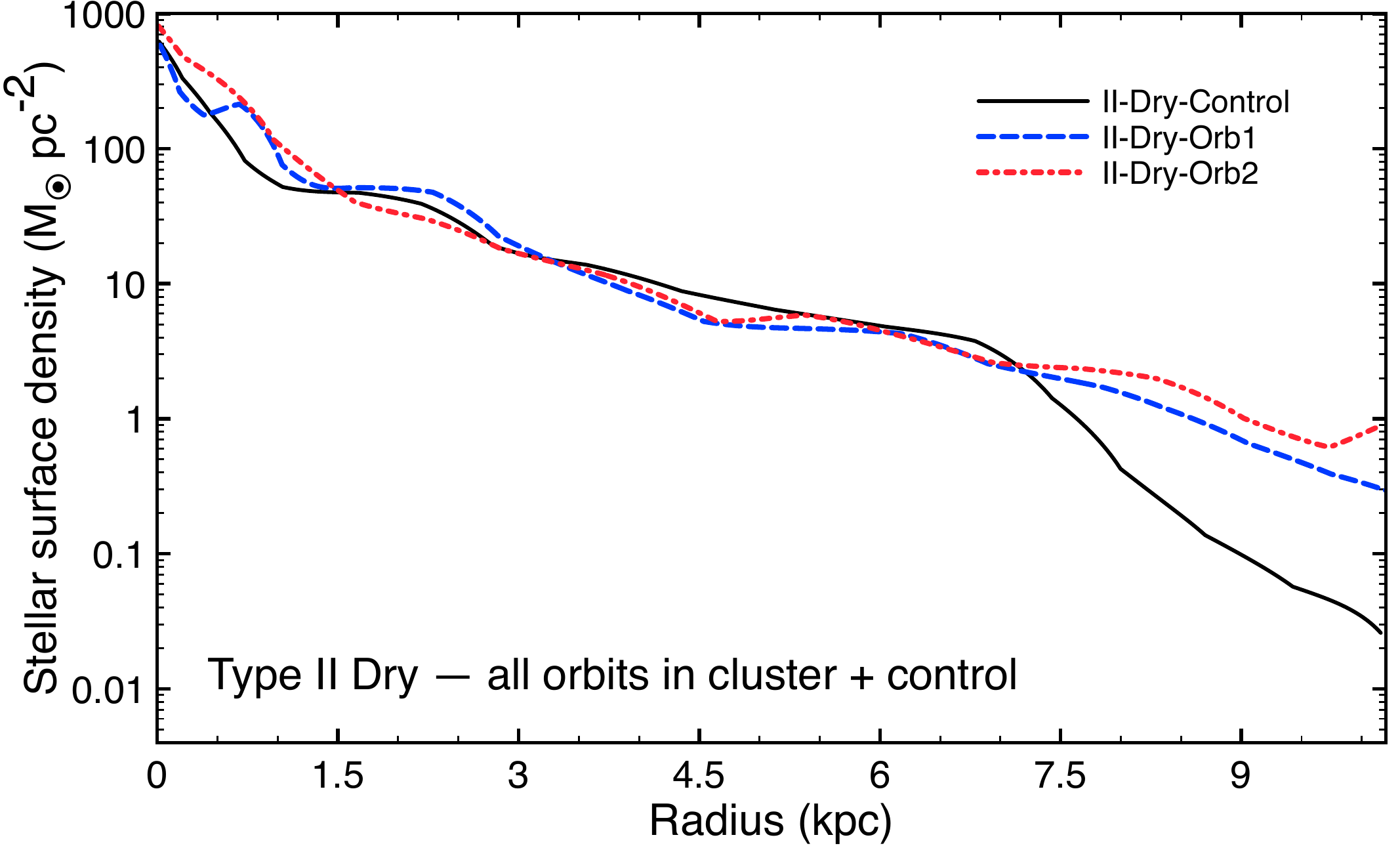}
\caption{Evolution of the Type~II 'dry' runs, for the 2 orbits simulated in the cluster, at $t=1.5$ (compared to the isolated control run shown with a solid black line).}
\label{fig:dry_evo}
\end{figure}

The evolution of several Type~II profile runs over a timescale of 1.5\,Gyr (i.e. typically about at the first cluster crossing) is shown in Fig. \ref{fig:wet_evo} and Fig. \ref{fig:dry_evo} for 'wet' and 'dry' runs, respectively.

We quantify the presence of a break in the density profile at any instant using the following quantities. The exponential slope $r_1$ of the inner disc is fitted between 1.5\,kpc (to exclude the central bulge) and 6\,kpc, which is slightly below the initial break radius. The exponential slope $r_2$ of the outer disc is fitted between 7.5\,kpc, slightly beyond initial break radius, and 10\,kpc. Note that this fitting choice assumes that the break radius remains approximately constant throughout the simulation, which is supported both by visual inspection of  profiles in our sample, and by earlier findings from BEE07, where break radii are found to be stable even in isolated discs.

We derive the ratio of inner to outer scale-length at any time $t$, defined as $\Delta(t) = \Delta_t =  r_2(t)/r_1(t)$. Type~I discs correspond to $\Delta=1$, Type~II discs to $\Delta>1$ and Type~III discs to $\Delta<1$. Our Type~II initial conditions are measured to have $\Delta_0$=3.36. By construction, our Type~III initial conditions have $\Delta_0$=0.33.

\begin{table}
\centering
\begin{tabular}{lcccccc}
\hline\hline
\noalign{\vskip 1pt}
\vspace{-.3cm}\\

Run &$\hspace{0.3cm}$& $\Delta_0$ &$\hspace{0.2cm}$& $\Delta_{1.5}$ 
&$\hspace{0.2cm}$& $\Delta_{2.0}$ \\
\vspace{-.3cm}\\
 
\hline  
\vspace{-.2cm}\\

II-Wet-Control  &&  3.36    &&   2.8  && 2.5  \\
II-Wet-Orb1  &&   3.36   &&    1.2  && 1.1 \\
II-Wet-Orb2  &&   3.36   &&    0.9  && 1.0 \\
II-Wet-Orb3  &&   3.36   &&    1.3   \\
II-Wet-Orb4  &&   3.36   &&    1.7   \\
II-Wet-Orb5  &&   3.36  &&    1.6   \\
II-Wet-Orb6  &&   3.36   &&   1.1\smallskip\\

II-Dry-Control  &&   3.36 &&  2.7 && 2.5   \\
II-Dry-Orb1  &&   3.36   &&   1.6 && 1.7   \\
II-Dry-Orb2  &&   3.36   &&   1.8&&1.4\smallskip\\

III-Wet-Control  &&  0.33    &&   0.37    \\
III-Wet-Orb1  &&   0.33   &&  0.46     \\
III-Wet-Orb2  &&   0.33   &&  0.35\smallskip \\

I-Wet-Control &&  1.0    &&   1.07    \\
I-Wet-Orb1  &&   1.0   &&  1.12     \\
I-Wet-Orb2  &&   1.0   &&  0.93 \\

\vspace{-.3cm}\\

\hline
\noalign{\vskip 2pt}
\hline
\end{tabular}
\caption{Simulation results: ratio of the inner and outer exponential scale-lengths ($\Delta_t$) at $t=0$, $t=1.5$ and $t=2.0$\,Gyr.}
\label{tab:slopes}
\end{table}

\begin{table}
\centering
\label{tab:scale-length}
\begin{tabular}{lccccc}
\hline\hline
\noalign{\vskip 1pt}
\vspace{-.3cm}\\
Run & $r_{1-0}$ & $r_{1-1.5}$ &$\hspace{0.1cm}$& $r_{2-0}$ & $r_{2-1.5}$ 
\\
\vspace{-.3cm}\\
 
\hline  
\vspace{-.2cm}\\

II-Wet-Control&  2.79  & 2.83  && 0.83 & 1.04 \\
II-Wet-Orb1  &   2.79  & 2.61  && 0.83 & 2.37 \\
II-Wet-Orb2  &   2.79  & 2.57  && 0.83 & 2.87 \\
II-Wet-Orb3   &   2.79  & 2.91  && 0.83 & 2.54 \\
II-Wet-Orb4  &   2.79  & 2.63  && 0.83 & 2.39 \\
II-Wet-Orb5  &   2.79  & 2.52  && 0.83 & 2.68 \\
II-Wet-Orb6  &   2.79  & 2.65  && 0.83 & 2.39 \smallskip\\

II-Dry-Control&  2.79 & 2.87 &&  0.83 &  0.98 \\
II-Dry-Orb1  &   2.79 & 2.76 &&  0.83 &  1.91 \\
II-Dry-Orb2  &   2.79 & 2.42 &&  0.83 &  2.38  \smallskip\\

\vspace{-.3cm}\\
\hline
\noalign{\vskip 2pt}
\hline
\end{tabular}

\caption{Evolution of inner and outer slopes for initially Type~II profiles. $r_{1-0}$ and $r_{1-1.5}$ are the inner exponential scale-lengths at 0 and 1.5 Gyr, measured between 1.5 and 6 kpc (to exclude the bulge and up to ~90\% of the break radius). $r_{2-0}$ and $r_{2-1.5}$ are the outer scale-lengths at the same epochs, measured between 7.5 and 10 kpc.}
\end{table}

\begin{figure}[tbp]
\includegraphics[width=0.46\textwidth]{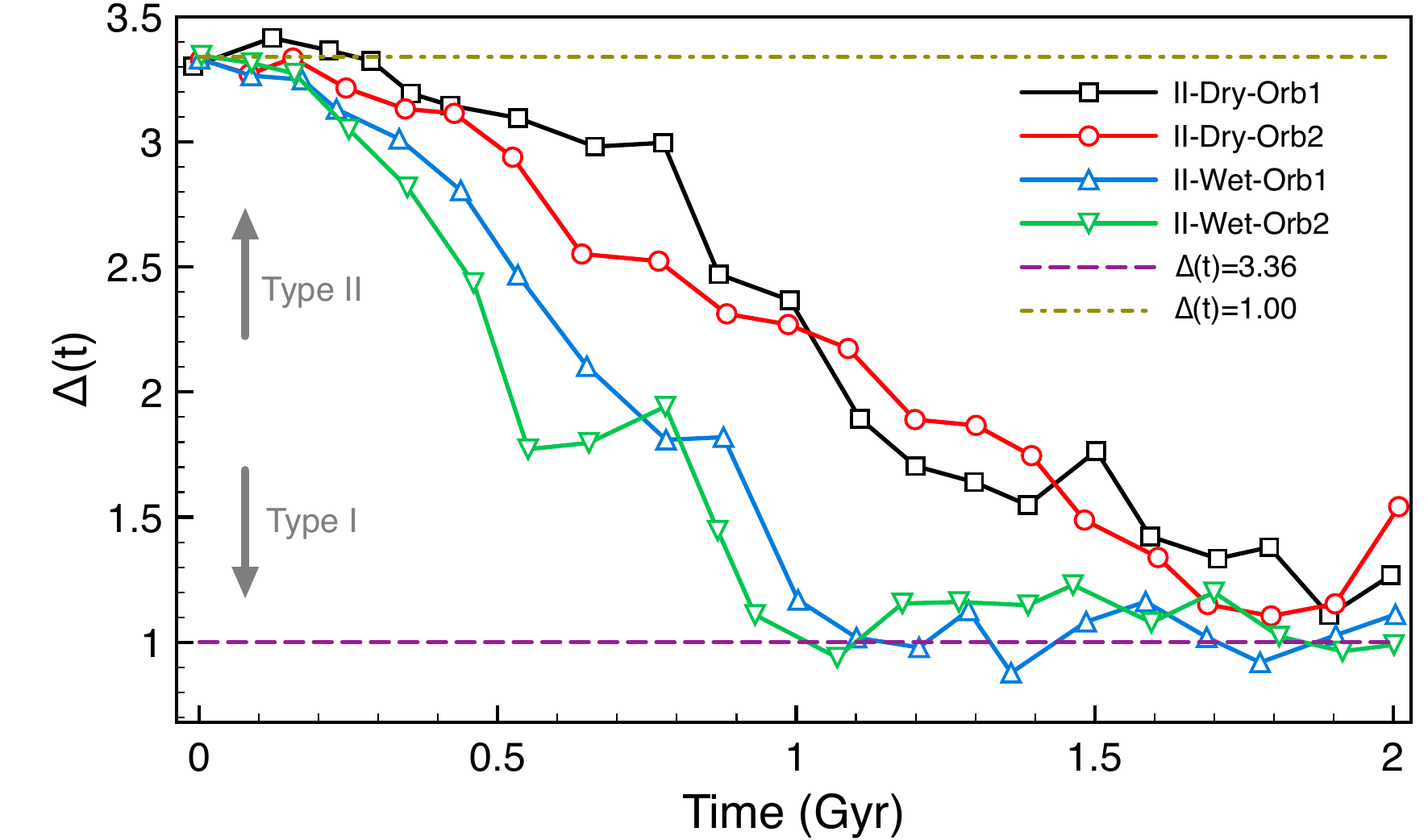}
\caption{Evolution of Type II galaxies: Time evolution of the ratio of inner to outer scale-length ($\Delta_t$) for 'wet' and 'dry' runs on two orbits.} 
\label{fig:Delta}
\end{figure}

Results on \(\Delta_0\) and \(\Delta_{1.5}\) for each simulation, and \(\Delta_{2.0}\) for the longer runs, are listed in Table~\ref{tab:slopes}. Simulations starting with Type~II discs at \(\Delta_0 = 3.0\) typically end up at \(\Delta_{1.5} \simeq 1\text{--}1.5\), implying that Type~II breaks are (mostly) erased. This holds true for both 'wet' and 'dry' models, i.e. either with or without star formation. Figure~\ref{fig:Delta} shows the time evolution of \(\Delta(t)\) for the two orbits (Orb1 and Orb2), which we ran for the longest duration, for both 'dry' and 'wet' runs. Note that in Fig. \ref{fig:Delta}, $\Delta=1$ corresponds to a pure Type~I exponential disc, values substantially larger than unity indicate a Type~II break, and our Type~II initial conditions have $\Delta_0=3.36$. To better understand whether the evolution in \(\Delta(t)\) is driven by changes in the inner or outer scale-lengths, we report in Table~\ref{tab:scale-length} the time evolution of \(r_1(t)\) and \(r_2(t)\) for Type~II discs. This reveals that while the inner scale-length remains nearly constant, the outer scale-length changes significantly, confirming that the disc evolution predominantly occurs in the outer regions. Note that the evolution of the orbital structure explains the gradual weakening of the break in the 'dry' cases. We attribute the faster evolution in 'wet' cases to the triggering of extra star formation in the outer disc by the cluster tidal field, whereas in 'dry' runs, the disappearance of disc breaks occurs more slowly, despite a similar rate of decrease in \(\Delta(t)\) by about a factor of two.

\subsection{Stellar motions and dynamical processes}
We here investigate here whether dynamical mechanisms acting on pre-existing stars, initially forming a Type\,II disc, can contribute to erase of the radial break into a Type~I profile.

\subsubsection{\label{evolutionstellarorbits}Evolution of stellar orbits}

\begin{figure*}
\sidecaption
\includegraphics[width=0.7\textwidth]{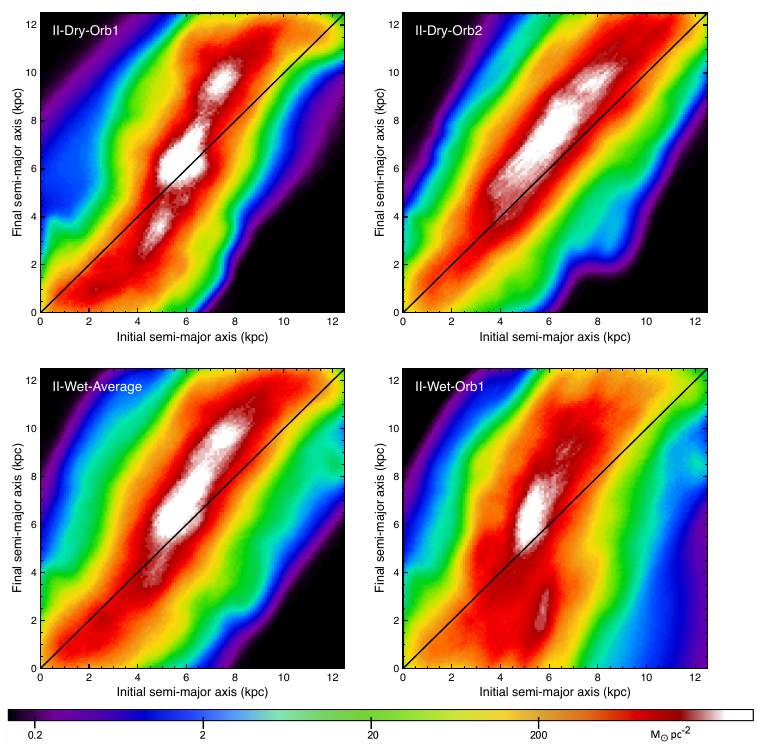}
\caption{Mass-weighted distribution of disc stars as a function of their initial and final semi-major axes at $t = 1.5$ Gyr. Top panels: results for the II-Dry-Orb1 (left) and II-Dry-Orb2 (right) simulations. Bottom panels: mass-weighted average over all 'wet' simulations (left) and a representative 'wet' case on a direct orbit (right). The solid black line marks the 1:1 relation (i.e. no change in semi-major axis). Note that this is not a particle-based plot but a binned, mass-weighted representation. The colour values (in M$_\sun$,pc$^{-2}$) indicate the stellar mass redistributed from a given initial to a final semi-major axis bin. This binning method can yield higher values at intermediate radii than in the central region, due to differing amounts of radial migration and local mass distribution.}
\label{sma-cases}
\end{figure*}

We examine how the evolution of the phase-space distribution of stars pre-existing to the galaxy-cluster interaction, and forming the initial Type~II disc, may explain or contribute to the evolution towards a Type~I profile. For this, we use two-dimensional histograms of stellar mass as a function of each star's initial and final semi-major axis. This representation provides a straightforward way to visualise radial migration: stars remaining on similar orbits lie along the 1:1 line, while net inflow or outwards migration appears as asymmetries relative to this line. Figure~\ref{sma-cases} displays the distribution for two representative 'dry' simulations (top panels; one on a direct orbit and one on a retrograde orbit), and presents in the bottom panels the mass-weighted average for all 'wet' simulations (left) and a representative 'wet' case on a direct orbit (right). This approach, inspired by works such as \citet{SellwoodBinney2002} and \citet{Roskar2008}, allows a compact visualisation of radial changes in orbital structure, which are a key mechanism behind the reshaping of disc profiles. While this projection focuses on changes in semi-major axis (as a proxy for orbital energy), it does not directly reflect variations in eccentricity or inclination. These other components of orbital structure are discussed separately below in the context of orbit heating and eccentricity changes.

The individual II-Dry-Orb1 case (top left panel of Fig.\ref{sma-cases}) shows that semi-major axes are decreasing in the inner regions, as expected for the direct interaction of a rotating disc with an external potential well: material inside the corotation loses angular momentum and flows towards the disc centre \citep[e.g.,][]{combesgerin}. Indeed, in this simulation, the estimated corotation radius -- defined as the radius where the angular velocity of disc material matches that of the external tidal perturbation -- lies at about 5\,kpc when the galaxy reaches its orbital pericentre, based on the galaxy rotation curve and orbital motion. Instead, material outside the corotation gains angular momentum through gravitational torques, and our measurements do show an increase in the semi-major axes of stars in the outer disc, with this transition occurring at radii somewhat smaller than the initial break of the Type\,II profile. The individual II-Dry-Orb2 case does not show any central inflow, and indeed such a retrograde case is not expected to experience significant gravitational torquing \cite{Quinn1993}. Nevertheless, the semi-major axes in the middle-to-outer disc still increase in a manner similar to the direct orbit. We attribute this increase in the semi-major axes to the increase in the eccentricity of stellar orbits. To estimate the eccentricity for each stellar particle, we assume that the enclosed mass acts as a central point mass (or a spherically distributed mass) and neglect the contribution from mass at larger radii. As a result, we find that the average eccentricity of the stellar orbits increases by a factor of 1.32 for stars initially located between 3 and 8 kpc from the centre, between $t=0$ and $t=1.5$\,Gyr. This increase in eccentricity explains the corresponding increase in the semi-major axes of stellar orbits. Hence, the observed larger semi-major axes can be attributed to the elongation of the orbits. This implies that stars initially located just beyond the radius where the radial profile breaks can contribute to increasing the density of the outer disc, at least for a portion of their orbit. As a result, this process could weaken or even erase the initial down-bending break in the radial profile.

Radial diffusion through a change in orbital structure is also present in 'wet' runs (bottom left panel of Fig.\ref{sma-cases}), with a similar increase factor in the outer disc as in the 'dry' runs, particularly around the initial break radius of the Type\,II profile. While we average over all 'wet' simulations, potentially masking individual variations, the representative example shown in the bottom right panel ('wet' Orb1) supports our interpretation: it clearly indicates that, beyond orbital restructuring, an additional mechanism, likely linked to the presence of gas, contributes to the accelerated evolution of the radial profile. This aligns with the faster disappearance of the break observed in 'wet' runs (Fig. \ref{fig:Delta}; see Sect. \ref{sc:ksrel}).

\subsubsection{Evolution of the disc vertical structure}

\begin{figure}[ht!]
\centering
\includegraphics[width=0.47\textwidth]{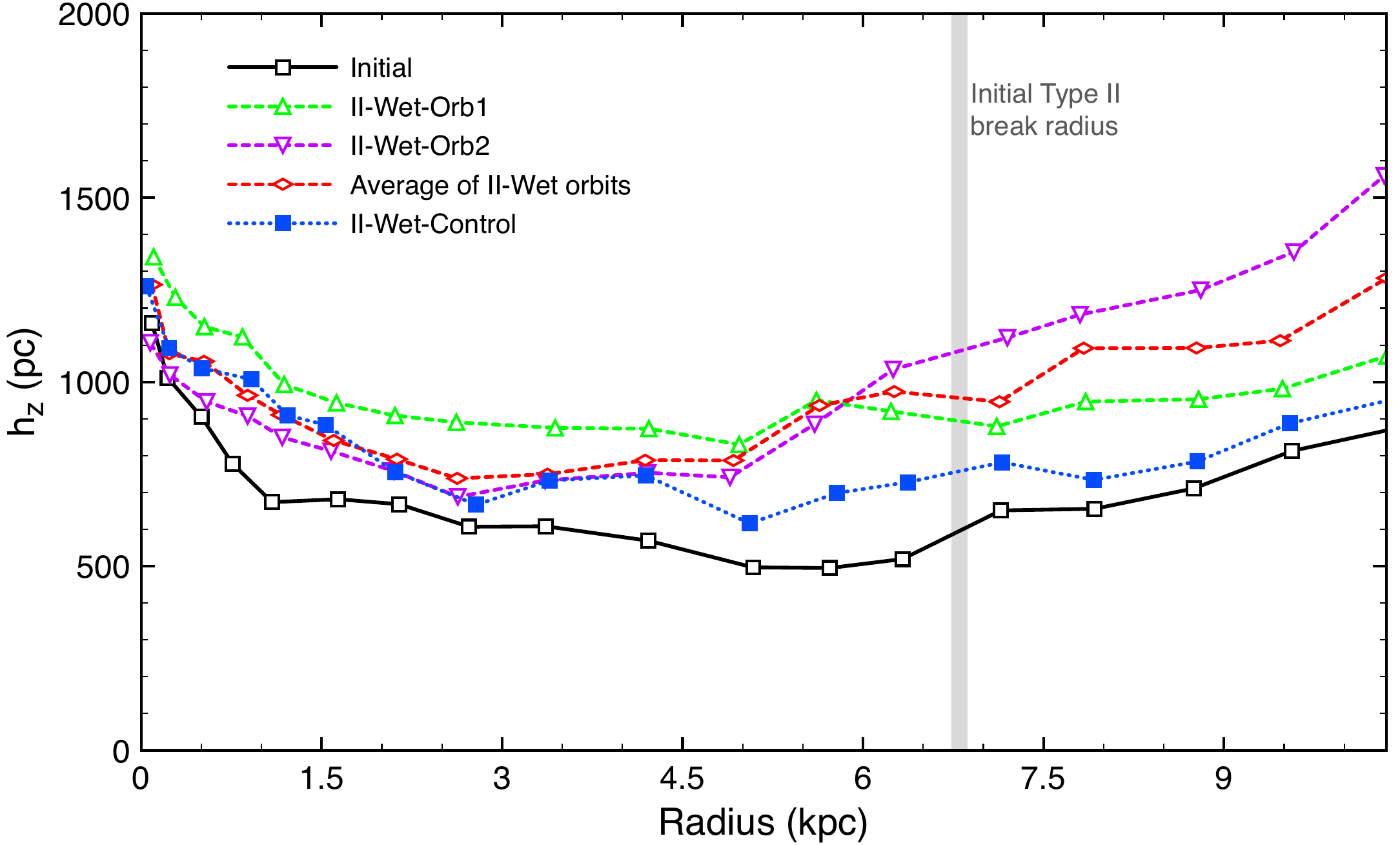}
\caption{Vertical scale-height of stellar discs as a function of radius in the initial conditions and several Type~II 'wet' runs at $t$=1.5\,Gyr. The scaleheight h$_z$ is fitted assuming an sech$^2$($z$/h$_z$) vertical profile.}
\label{fig:Zprof}
\end{figure}

We also examine the impact of the cluster on the evolution of disc vertical profiles Type~II disc, assuming vertical density profiles of the form sech$^2$($z$/h$_z$), where $z$ is the elevation above the disc mid-plane and h$_z$ the vertical scaleheight. Representative cases and the average evolution over all orbits are displayed in Fig.~\ref{fig:Zprof}. The initial disc shows thickening in the inner region, likely to result from bulge contamination, and moderate thickening in the outer regions, which here likely results from the formation of a thick discs by internal instabilities \citep*[e.g.,][]{bournaud09} but may also results from interactions and mergers with minor companions \citep[e.g.,][]{DiMatteo2019}.

In isolation, the disc vertical scale-height secularly increases (by no more than 10-20\% per Gyr) especially in the bar and spiral arm regions where vertical resonances can be triggered, especially near the bar extremity and corotation resonance, as described by \cite{bar-thickening} and \cite{bar-thickening2}.

When interacting with the cluster, the vertical evolution of the disc shows substantial differences. On average, it is similar to the isolated case in the inner $\sim$ 4--5\,kpc, i.e. in the dense main disc regions, and increases up to to a factor 1.5-2 at larger radii. Individual cases however show substantial differences. Direct orbits (such as the II-Wet-Orb1 case displayed in Fig~\ref{fig:Zprof}) show weaker-than-average thickening in the outer (>\,6\,kpc) regions but stronger-than-average in the inner ones. This indicates that additional processes are at play in the inner region for direct orbits, likely due to gravitational torquing and the removal of angular momentum from the inner disc by the co-rotating cluster field. In contrast, retrograde orbits (such as II-Wet-Orb2 on Fig~\ref{fig:Zprof}) show stronger-than-average thickening in the outer (>\,6\,kpc) but weaker-than-average in the inner ones: this indicates that the thickening process is dominated by tidal stirring \citep{Kauffmann1993}, to which low-density outer regions are most sensitive, with little contribution of gravitational torquing. This appears similar to the known fact that gravitational torquing is stronger on prograde orbits than on retrograde orbits in the case of galaxy-galaxy interactions (leading in particular to stronger nuclear starbursts and the formation of longer tidal tails, e.g., \citet{B04}, \citet{galmer}, \citet{tjcox}. These results are consistent with the conclusions drawn from the 2-D radial profile analysis (see Sec.\ref{evolutionstellarorbits}), where signatures of torquing and associated resonance dominate the distribution of stellar orbits on direct cases, while retrograde orbits appear dominated by outer disc stirring.

\subsection{Evolution by star formation}\label{subsec:sfr}

We now explore how star formation itself contributes to the morphological transformation of galactic discs, particularly under the influence of a cluster tidal field. We first analyse the star formation behaviour in our simulations via the Kennicutt-Schmidt relation, before interpreting these trends using a simple toy model.

\subsubsection{The Kennicutt-Schmidt relation}\label{sc:ksrel}

\begin{figure}[ht!]
\includegraphics[width=0.47\textwidth]{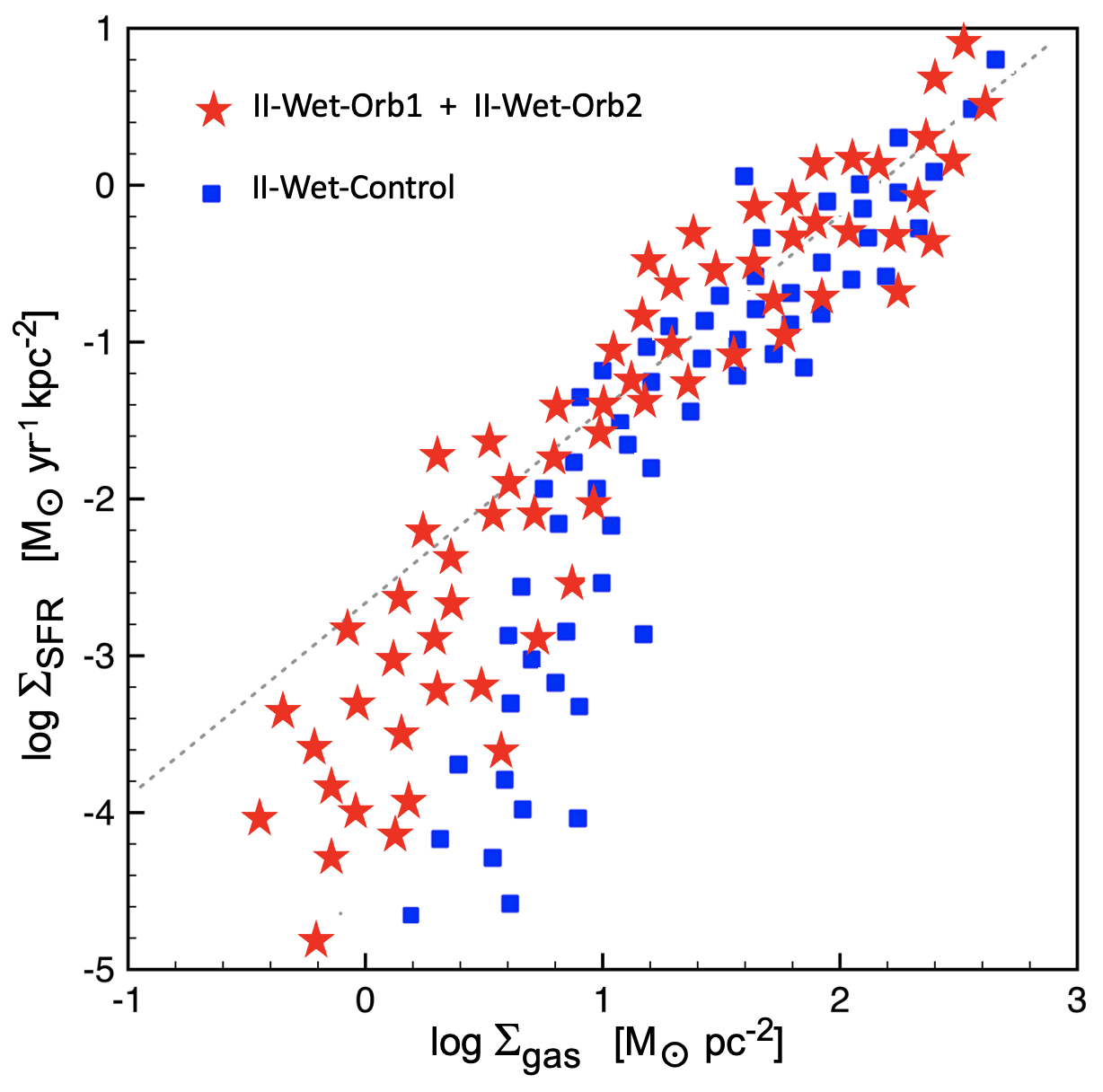}
\caption{Relation between gas surface density and star formation rate surface density, measured with the disc seen face-on and over circular regions of radius 500\,pc. The centre of each regions is selected randomly outside the central 2\,kpc (we do not aim at studying the bulge region and contamination through it). The dashed black line is the relation defined for nearby disc galaxies by \citet{leroy}. The II-Wet-Control run is shown with blue squares and the average or II-Wet-Orb1 and II-Wet-Orb2 with red squares, no meaningful difference is find between the two runs, or with other II-Wet runs.} 
\label{fig:ks}
\end{figure}

We begin by examining the star formation behaviour in the control simulation, which evolves in isolation without the influence of a cluster potential. This provides a reference framework to assess the impact of environmental effects. The isolated disc follows the expected Kennicutt-Schmidt relation, with a nearly constant gas consumption timescale for gas surface densities above $\sim$10\,M$_\sun$\,pc$^{-2}$ -- well within the inner disc, typically at $\lesssim 0.5\,R_{\rm break}$ -- and shows a marked decline in star formation efficiency below this threshold. This downturn corresponds to a 'break' similar to that observed in nearby galaxies \citep[e.g.,][]{Kennicutt1998, leroy}. We then study how this relation is modified in several 'wet' simulations that include the cluster potential, focusing on its effect on star formation in an initially Type~II disc (Fig.\ref{fig:ks}). Note that in Fig.\ref{fig:ks}, the II-Wet-Control has a star formation efficiency that drop below surfaces densities about 10\,M$_\sun$\,pc$^{-2}$, when the same galaxy interacts with the cluster tidal field, the efficiency drop is pushed towards lower surface densities, and is shallower.

Galaxies interacting with the cluster potential well display different properties. For gas surface densities above $\sim$10\,M$_\sun$\,yr$^{-1}$, the star formation efficiency is about similar to the isolated one -- with on average a slightly higher efficiency, by a factor 1.2. The key difference is below the 10\,M$_\sun$\,yr$^{-1}$ density threshold. The star formation efficiency does still drop, but in a more moderate and slower fashion. At gas surface densities around 3\,M$_\sun$\,pc$^{-2}$, 'wet' simulations undergoing pericentre passage within the cluster potential show star formation efficiencies 10--50 times higher than in the isolated case. Conversely, similar star formation rate surface densities can be found at gas surface densities ten times lower when the cluster potential is present. We will provide a theoretical interpretation for this below. In any case, the figures measured here mean that the outer disc beyond the initial break radius experiences extra star formation at a surface density rate about 10$^{-3}$\,M$_\sun$\,yr$^{-1}$\,kpc$^{-2}$. When integrated over a 300\,Myr period around the pericentre this corresponds to forming an extra $\sim$1\,M$_\sun$\,pc$^{-2}$, which is the order of magnitude needed to fill the drop in the density profile in the few kpc beyond the initial Type~II profile. This shows that star formation triggered by the galaxy-cluster interaction in the low-density outer disc provides a stellar density close to what is required to erase the disc down-bending break. Since the order of magnitude remains similar to the one provided by the change in orbital structure (Sect.~\ref{evolutionstellarorbits} above) this explains why the dissolution of Type~II down-bending breaks is faster in 'wet' than in 'dry' cases, but not by an order of magnitude, rather by a factor $\sim$2 (Fig.~\ref{fig:Delta}).

While directly verifying that the stars populating the outskirts have been recently formed would be ideal, this remains challenging in practice. The difficulty arises from the combined effects of newly formed stars and stellar migration, both contributing to the outer disc population and making it difficult to disentangle their respective roles based solely on the final stellar distribution. Figure~\ref{fig:Delta} clearly shows that although 'dry' runs gradually smooth out Type II profiles, 'wet' runs achieve this more rapidly. This accelerated evolution in 'wet' runs can be attributed primarily to the triggering of additional star formation in the outer disc induced by the cluster tidal field. Moreover, the presence of gas and star formation in 'wet' runs likely influences the stellar orbits and dynamical evolution, further contributing to the faster erosion of Type II breaks. Thus, the interplay between in-situ star formation and radial migration in the outer disc regions adds complexity to interpreting the stellar population data and underscores the necessity for more detailed analyses to fully understand the mechanisms responsible for the weakening of Type II disc breaks.

\subsubsection{A toy model for star formation triggering}

We estimate the influence of cluster tidal forces -- modelled after a Perseus-like cluster -- on a typical disc galaxy, focusing on their capacity to induce turbulence and gas compression that may trigger star formation. We place the galaxy at $r = R_{\text{c,1/2}} = 200$\,kpc, the cluster half-mass radius, corresponding to a cluster mass $M_{\text{c,1/2}} = 1.3 \times 10^{15}$\,M$_\odot$.

To assess the gravitational stability of the gas, we evaluate the Jeans length \citep{Binney1987}:

\begin{equation}
    \lambda_J = \sqrt{\frac{\pi c_s^2}{G \rho_{\text{gas}}}},
\end{equation}

assuming $c_s = 10$\,km\,s$^{-1}$ (appropriate for gas at $10^4$\,K) and an average gas density $\rho_{\text{gas}} = 0.076$\,M$_\odot$\,pc$^{-3}$, representative of the outer disc (7–12\,kpc). These values yield $\lambda_J \approx 980$\,pc. The associated free-fall time \citep{Binney1987} is:

\begin{equation}
    t_{\text{ff}} = \sqrt{\frac{3 \pi}{32 G \rho_{\text{gas}}}} \approx 28\,\text{Myr}.
\end{equation}

To quantify the dynamical effect of tides, we approximate the tidal acceleration over a region of size $\lambda_J$ as:

\begin{equation}
    a_{\text{tidal}} \approx \frac{2 G M_{\text{c,1/2}} \lambda_J}{R_{\text{c,1/2}}^3}.
\end{equation}

This results in a displacement $d \approx 620$\,pc over one free-fall time. Consequently, the gas experiences a contraction from $\lambda_J$ to $\lambda_J - d \approx 360$\,pc. This corresponds to a one-dimensional compression factor of nearly 3. In three dimensions, assuming isotropic compressive forces, the density could increase by more than an order of magnitude.

In addition to compression, the same tidal acceleration induces turbulent motions. Over $t_{\text{ff}}$, the typical velocity reached is $v_{\text{induced}} \approx 40$\,km\,s$^{-1}$. For atomic gas at $10^4$\,K, this corresponds to a Mach number $\mathcal{M} \approx 4$, firmly in the supersonic regime. Supersonic turbulence efficiently drives local density fluctuations and is expected to decay over a free-fall time \citep[e.g.][]{maclow99}, thereby promoting gas collapse and star formation.

Tidal forces also inject turbulence into both solenoidal and compressive modes. As shown by \citet{federrath}, gravitational interactions naturally favour solenoidal driving ($\sim 2/3$), yet compressive modes can dominate locally in regions of converging flows. In particular, the compressive component is enhanced where the trace of the tidal tensor is negative \citep{Renaud2014}, leading to focused compressive regions that further stimulate star formation.

These combined effects can significantly alter the outer disc structure. In our simulations, the average gas surface density between 7–12\,kpc is $\sim2$\,M$_\odot$\,pc$^{-2}$, corresponding to a star formation rate surface density $\Sigma_{\text{SFR}} \approx 10^{-2.4}$\,M$_\odot$\,yr$^{-1}$\,kpc$^{-2}$. Over a 200\,Myr interaction (corresponding to one cluster crossing time at 250\,km\,s$^{-1}$ over 500\,kpc), this leads to the formation of $\sim1$\,M$_\odot$\,pc$^{-2}$ of new stars. Such a contribution is sufficient to smooth out or erase a pre-existing Type~II (down-bending) disc break, as shown in Fig.~\ref{fig:ini_prof}.

While observational confirmation through stellar population gradients is complicated by low signal-to-noise ratios at large radii, the theoretical mechanism presented here provides a consistent explanation for the morphological transformation of disc galaxies in clusters. We therefore find that tidal turbulence and compression can act on short timescales, in complement to slower processes such as harassment or gas starvation, to reshape galactic discs. The enhanced turbulence effectively lowers the threshold for collapse, sustaining star formation throughout the interaction. In gas-rich systems, this may erase Type~II breaks entirely and even produce the extended light profiles characteristic of Type~I or Type~III discs.

In summary, this simplified model illustrates how cluster tides can generate supersonic turbulence, compress the gas, and trigger star formation. These processes act efficiently within one crossing time, providing a plausible physical mechanism for the increased frequency of Type~I and Type~III profiles observed in dense environments.

\section{\label{sc:results13}Results: Evolution of Type~III (up-bending break) and Type~I (pure exponential) profiles in/near a galaxy cluster}

\subsection{Survival of Type~III profiles}

Our simulations show that Type~III profiles are generally preserved during the infall of galaxies into a cluster environment. The inner and outer scale lengths (\(r_{\text{1}}\), \(r_{\text{2}}\)) remain essentially stable throughout the interaction. We do observe a slightly higher \(\Delta_{1.5}\) in the III-Wet-Orb1 simulation compared to III-Wet-Orb2, suggesting that prograde orbits may have a somewhat stronger impact on the outer scale length \(r_{\text{2}}\) than retrograde orbits. However, these variations in \(\Delta\) remain far less significant than those observed in Type~II discs.

Although the number of Type~III systems in our sample is limited, they consistently retain their up-bending profiles regardless of orbital configuration. The modest variations in \(\Delta\) we observe are primarily associated with mild changes in the outer disc, which may become slightly more extended. This could potentially be linked, as for Type~II galaxies, to a small amount of additional star formation occurring in the outer regions during the interaction, further reinforcing the anti-truncation.

We also note that, in contrast to Type~II systems, the outer gas in Type~III galaxies lies within a relatively dense stellar disc. Such an embedding has been shown to increase the star formation efficiency of a gas disc \citep{martig09, martig13, genzel14}, which could contribute to the robustness of Type~III profiles in a variety of environments.

\subsection{Survival of Type~I profiles and limits on the formation of Type~III breaks}

We also find that Type~I profiles are relatively unaffected for the modelled disc galaxy (Table~\ref{tab:slopes}). Star formation effects appear not to be sufficient to dominate over the disturbance of orbits, so no robust up-bending break appears. The absence of a clear transition from Type~I to Type~III profiles in our simulations likely reflects both the initial gas distributions and the processes modelled. Our discs are initially gas-rich with smooth radial profiles, lacking central gas deficits or extended outer rings often observed in real galaxies \citep[e.g.,][]{Leroy2008}. In such conditions, star formation triggered by tidal effects occurs over a broad radial range, producing relatively uniform stellar profiles consistent with Type~I rather than building the outer excess characteristic of Type~III breaks.

\section{\label{sc:Discussion}Discussion}

Our simulations provide new insights into the effects of cluster environments on disc galaxy profiles, and help interpret the observational findings from the Euclid ERO of the Perseus cluster (Paper~I). Specifically, we highlight how cluster tidal forces interact with star formation processes and orbital dynamics to shape the evolution of the radial and vertical mass/light profiles, offering explanations for the observed distribution of Type~I, Type~II and Type~III profiles in Perseus.

\subsection{Evolution of Type~II discs versus observations}\label{subsc:DiscussionTypeII}

For Type~II profiles, our simulations reveal that tidal interactions in a cluster environment generally lead to a rapid reduction in the prominence of breaks, especially in 'wet' cases where galaxies contain a substantial amount of gas and star formation is triggered in the outer regions. This star formation smooths out the down-bending break faster than dynamical effects alone would do. In 'dry' cases, without gas and star formation, the break is also smoothed out within less than 2 Gyr, mostly due to changes in orbital structure. This aligns with the findings in Paper~I and in \citet{Erwin2012}, which show that Type~II profiles are scarce in the Perseus and Virgo clusters, respectively.

Our results are consistent with, and complementary to, previous N-body simulation studies that have explored the evolution of disc breaks through internal or accretion processes. For example, \citet{Minchev2012} and \citet{Clarke2017} demonstrated that radial migration, bar and spiral instabilities, and minor mergers can transform Type~II profiles into Type~I or Type~III profiles in isolated galaxies. \citet{Pfeffer2022}, using cosmological N-body simulations, further showed that the diversity of disc profiles can be explained by a combination of migration, accretion, and star formation history. However, these studies mainly focus on galaxies in isolation or in cosmological volumes where the external tidal field is relatively weak or averaged out. Our hydrodynamical simulations show that in a dense cluster, strong and persistent tidal forces can accelerate the erasure of Type~II breaks, providing a direct explanation for their observed scarcity in rich clusters.

These simulation results are also in excellent agreement with recent observational studies. Both \citet{Pranger2017} and \citet{Raj2019} present systematic analyses of disc profile types in clusters, confirming that the fraction of Type~II discs decreases significantly in dense environments, while Type~I and Type~III profiles become more prevalent. This supports the view that strong environmental mechanisms in clusters are responsible for erasing or transforming down-bending breaks.

Several other observational studies have shown that disc profile types can be influenced by environment. For instance, \citet{ErwinBeckmanPohlen2005} and \citet{Gutierrez2011} found that down-bending (Type~II) breaks are sensitive to dense environments, suggesting an external origin in some cases. Similarly, \citet{pohlen06} reported evidence for tidal influences on Type~II discs. Complementary, \citet{comeron2012} observed that Type~I profiles in clusters often exhibit extended recent star formation, consistent with our simulations where star formation is triggered in the low-density outskirts, contributing to the smoothing of Type~II breaks.

There are still a few Type~II galaxies in the cores of the Perseus and Virgo clusters. They could correspond to galaxies that entered the cluster only recently, or to rare orbital configurations favourable to the survival of the breaks, not spanned by our simulation sample. Although we do not quantify precise destruction rates, comparisons between prograde and retrograde orbits in our simulations show that both types of galaxies experience similar tidal friction effects. This demonstrates that, regardless of their orbital direction, galaxies do undergo structural changes when interacting with the cluster environment. However, more orbits would be needed to make detailed comparisons. Additionally, some Type~II galaxies may result from ram-pressure stripping of the outer gas disc, preventing the triggering of star formation. It is also possible that particularly strong bars help maintain a Type~II profile by compensating for the cluster tidal effects, since bars can both trigger the rise of Type~II profiles and limit radial migration through their resonances \citep{Debattista2006, athanassoula2002}. However, for moderate bars (present in our sample) the effect appears insufficient to compensate for the tidal effects exerted by the cluster.

\subsection{Evolution of Type~I and III discs versus observations}

In our simulations, neither clear transitions from Type~I to Type~III discs nor the emergence of new up-bending breaks are observed. Rather, both Type~I and Type~III galaxies tend to preserve their initial structural profiles throughout their cluster passage. For one specific orbital configuration with a gas-rich disc, the structural stability of Type~III profiles is accompanied by a slight outward increase in the outer scale length \(r_2\), while the inner scale length \(r_1\) remains largely unchanged. This moderate change could reflect mild tidally-triggered star formation in the outskirts rather than significant stellar redistribution. Variations in \(\Delta_{1.5}\) are small compared to those observed in Type~II discs, underscoring the relative robustness of Type~III profiles.

A modest orbital dependence is noted, with slightly larger \(\Delta_{1.5}\) increases for prograde orbits (e.g., III-Wet-Orb1) compared to retrograde orbits (e.g., III-Wet-Orb2), implying that tidal torques aligned with disc rotation may more effectively stimulate outer-disc activity. Nonetheless, these differences remain limited and do not fundamentally change the overall profile classification.

The persistence of both Type~I and Type~III profiles in our models aligns well with the trends reported in Paper~I, which found that Type~I and Type~III discs dominate the morphological distribution of cluster galaxies. Observationally, outer \ion{H}{i} gas components in Type~III galaxies are often associated with extended star-forming rings or disturbed outskirts, especially in galaxies on cluster-peripheral or eccentric orbits \citep[e.g.,][]{Leroy2008}. These outer gas reservoirs can sustain star formation beyond the up-bending break, contributing to the maintenance or mild enhancement of Type~III profiles under tidal perturbations.

However, the absence of clear Type~III formation from Type~I in our simulations also points to key limitations. Our models cover only a restricted set of initial conditions and orbital parameters, all based on single, idealised cluster passage. Real cluster environments are more complex: gas morphologies often include central \ion{H}{i} holes and outer gas rings, which observationally favour central quenching and ring-like outer star formation \citep{Leroy2008}. Such gas distributions are not captured in our initial conditions, which assume smooth, fully gas-rich discs. Additionally, processes like ram-pressure stripping or repeated interactions are not included, yet they can strongly affect gas availability and the structural evolution of outer discs. Their inclusion could limit the formation of Type~III profiles, especially in cluster cores, or conversely, enhance them in the outskirts through minor mergers or accretion events. \citep{Younger2007, Pfeffer2022}

This is further supported by colour profile analyses in Paper~I, where the majority of Type~III galaxies showed flat or red gradients -- indicative of no specific assymetric star formation -- while a minority exhibited U-shaped profiles suggestive of stellar migration or accretion. The latter processes are not present in our simulations and could help explain the diversity observed. Similarly, N-body simulations by \citet{Minchev2012}, \citet{Clarke2017}, and \citet{Pfeffer2022} have shown that radial migration, secular evolution, and minor mergers can all contribute to Type~III-like structures, especially in older stellar populations.

For Type~I profiles, our simulations confirm their structural resilience under tidal perturbations, in agreement with previous work \citep[e.g.,][]{Elmegreen2006, Martig2009, Pfeffer2022}. The symmetric mass distribution and lack of significant gas asymmetries in our models likely contribute to their robustness. This may explain the high prevalence of Type~I profiles near the cluster core observed in Paper~I, where galaxies on circular orbits experience strong but relatively stable tidal fields. Moreover, as discussed in Section \ref{subsc:DiscussionTypeII}, a significant fraction of Type~I profiles could originate from the secular evolution of Type~II discs, providing an additional explanation for their abundance. However, this resilience might break down under more complex dynamical conditions or environmental processes not explored in our current simulations.

In summary, while our results reinforce the idea that Type~I and Type~III profiles are structurally stable during a single cluster passage, the limited diversity of initial conditions, orbital histories, and environmental processes in our models restricts the generality of these conclusions. Further studies incorporating more realistic gas distributions, multiple interactions, and broader environmental physics are required to fully understand the formation and diversity of Type~III breaks in cluster environments.

\subsection{Disc breaks along cosmic time and environments}\label{sc:environments}

Interestingly, recent JWST observations \citep{jwst-breaks} suggest that Type~II discs are relatively common at high redshift ($z\sim2$), an epoch where violent disc instabilities can rapidly form double-exponential, Type~II profiles \citep[e.g.,][]{BEE07}. In our simulations, we do not model the formation of these profiles, but instead start from galaxies that already exhibit Type~II breaks, likely shaped by such early internal processes. We then explore how these profiles evolve as galaxies interact with a cluster environment. Our results show that these initially well-defined Type~II breaks are often weakened or erased due to tidal perturbations from the cluster potential. This suggests that the structural transformation of disc galaxies in clusters can be primarily driven by gravitational processes -- internal ones (e.g., bars, resonances) forming the breaks, and external tidal effects eroding them. In this framework, hydrodynamical effects such as ram pressure stripping, or star formation thresholds, seem to be not required to explain the observed disappearance of Type~II profiles in dense environments, although they may still contribute in some cases. Thus, the presence of Type~II profiles at high redshift, their persistence in the field at low redshift, and their relative rarity in low-redshift clusters can be interpreted as different stages of a common evolutionary pathway dominated by gravitational dynamics.

\section{\label{sc:Conclusions}Conclusion}

In this study, we conducted hydrodynamical simulations to investigate how cluster environments influence the structural evolution of the radial stellar density profiles of disc galaxies. The cluster environment was modelled using the basic parameters of the Perseus cluster, allowing direct comparison with recent Euclid ERO data in the Perseus field \citep{LF}. Using the AMR code RAMSES \citep{ramses}, we simulated galaxies with initial Type~I, Type~II, and Type~III radial profiles. Notably, the initial Type~II profiles represent disc structures formed through internal dynamical evolution at high redshift \citep[e.g.,][]{BEE07}. These model galaxies were evolved both in isolation and within a Perseus-like tidal field. By varying parameters such as gas fraction, star formation efficiency, and orbital inclination, our simulations were designed to analyse the combined — and, where possible, disentangled — effects of star formation and tidal forces on the density profiles of disc galaxies in cluster conditions. In particular, by comparing 'dry' (gas-free) and 'wet' (gas-rich) runs, we aimed to isolate the role of gas processes in the evolution of Type~II profiles. This controlled setup enabled systematic contrasts between isolated evolution and cluster-induced transformations, with specific relevance to the Perseus cluster as observed in Paper~I \citep{Mondelin2025}.

Our results demonstrate that tidal interactions affect both the pre-existing orbital structure of stellar discs and the star formation efficiency therein, leading to a rapid and near-complete smoothing of Type~II profile breaks across all simulated orbital configurations, with this process occurring more rapidly in gas-rich galaxies. This trend is consistent with the observed scarcity of Type~II profiles in the Euclid ERO Perseus field (see Paper~I), although some rare Type~II profiles persist in the cluster core. These may be stabilised by internal structures such as strong bars or enhanced by ram-pressure stripping, which are not included in our current models. While our simulations do not produce strong bars, future work could incorporate them to investigate their impact on Type~II profile longevity.

Both Type~I and Type~III profiles remain largely stable during a single cluster passage, with no clear transitions from Type~I to Type~III nor significant formation of new up-bending breaks. For Type~III discs, a modest outward increase in the outer scale length is observed in some prograde, gas-rich cases, likely driven by mild tidally-triggered star formation, whereas Type~I profiles exhibit strong structural resilience to tidal perturbations. These findings align with observational trends from Paper~I but are constrained by the limited range of initial conditions and orbital parameters explored, as well as the exclusion of complex environmental effects such as ram-pressure stripping and repeated interactions that may strongly influence disc evolution in real cluster environments.

Overall, our simulations support and extend observational findings from Perseus. They suggest that gravitational tidal effects of the cluster are the main driver behind the pronounced scarcity of Type~II breaks and the dominance of Type~I profiles. This occurs primarily through their impact on both stellar orbital structure and star formation efficiency in the outer disc regions. This scenario is consistent with a picture where disc breaks form early via internal dynamical instabilities and are subsequently eroded in cluster environments, without necessarily invoking additional mechanisms such as ram-pressure stripping or star formation thresholds. Nonetheless, processes not included in our models—such as bar evolution, galaxy-galaxy interactions, and minor mergers—probably play a role in shaping disc profiles in more complex ways. Future simulations incorporating these factors, alongside a broader range of initial galaxy properties, will be essential to refine our understanding of the interplay between internal processes and cluster-induced forces in shaping disc galaxy structure.

\begin{acknowledgements}
We are indebted Sandrine Codis for her suggestions and comments.
This work was motivated by the Euclid ERO Perseus observations, and we thank all the participants in this observational effort. We are indebted to Connor Stone for his help in measuring radial luminosity profiles in faint regimes. This project was provided with computer and storage resources by GENCI at TGCC thanks to the grant 2023-A0150402192 on the supercomputer Joliot Curie's ROME partition .
\end{acknowledgements}

\bibliographystyle{aa} 
\bibliography{aa54840-25}

\end{document}